# Silicon solar cells efficiency analysis. Doping type and level optimization

A.V. Sachenko, V.P. Kostylyov, M.V. Gerasymenko, R.M. Korkishko, N.R. Kulish, M.I. Slipchenko, I.O. Sokolovskyi, V.V. Chernenko

Abstract

The theoretical analysis of photovoltaic conversion efficiency of highly effective silicon solar cells (SC) is performed for n-type and p-type bases. The case is considered when the Shockley-Read-Hall recombination in the silicon bulk is determined by the deep level of Fe. It is shown that due to the asymmetry of the recombination parameters of this level the photovoltaic conversion efficiency is increasing in the SC with the n-type base and decreasing in the SC with the p-type base with the increase in doping. Two approximations for the band-to-band Auger recombination lifetime dependence on the base doping level are considered when performing the analysis.

The experimental results are presented for the key characteristics of the solar cells based on $\alpha - Si:H - n - Si$ heterojunctions with intrinsic thin layer (HIT). A comparison between the experimental and calculated values of the HIT cells characteristics is made. The surface recombination velocity and series resistance are determined from it with a complete coincidence of the experimental and calculated SC parameters' values.

## Introduction

When analysing dependence of the silicon solar cells (SC) on the base doping level, the assumption is commonly used that the deep levels close in $E_t$ energy to the middle of the band gap with close electron- and hole-capture cross sections are responsible for the Shockley-Read-Hall (SRH) recombination. Specifically, the Au level meets these criteria [1]. At the same time, the Shockley-Read-Hall recombination can be determined by the energy levels which do not agree with the middle of the band gap and electrons $\sigma_n$ and holes $\sigma_p$ cross-sections differ essentially. The level of Fe is one of such levels [2, 3]. According to [3] it is characterized by the values $E_c - E_t$ = 0.774 eV ($E_c$ is the conduction band edge), $\sigma_n$ = 5·10$^{-14}$ cm$^2$, $\sigma_p$ = 7·10$^{-17}$ cm$^2$. In this case the Shockley-Read-Hall lifetime value $\tau_{SRH}$ can significantly depend both on the SC base doping and on the excess electron-hole pairs' concentration $\Delta n$ in the base. These dependences can significantly affect the minority-carriers effective bulk lifetime $\tau_{eff}$ determined by the relation of different recombination mechanisms, including the Shockley-Read-Hall recombination, radiative recombination and band-to-band Auger recombination.

The analysis of the photovoltaic conversion efficiency $\eta$ is carried out in this research for highly effective silicon SC and HIT (heterojunction with intrinsic thin layer) SC, depending on the base type and doping level for the case when the Shockley-Read-Hall recombination lifetime is determined by the Fe level. Calculation results are compared to the experimental values of the photovoltaic conversion efficiency $\eta$, open-circuit voltage $V_{OC}$, fill-factor $FF$ and some other parameters of the SC based on $\alpha-Si:H-n-Si$ SC.

A simple approach is used in the analysis, which makes it possible to model characteristics of the SC produced on the crystalline base [4]. Its special feature is the fact that one of the main characteristics of the SC is the short circuit current density $J_{SC}$ is found experimentally and the remaining SC parameters are calculated. This essentially simplifies the analysis of the experimental results, which can serve as a basis for optimization of such characteristics of the HIT cells as the doping level of the base $N_d$ ($N_a$) under the given Shockley-Read-Hall lifetime $\tau_{SRH}$ value, surface recombination velocity $S_0$ and $S_d$ on the SC frontal and rear surfaces and series resistance $R_s$.

## 1. Analysis of $\tau_{SRH}$ dependence on the base doping level for the cases of the base area of p- and n-type

Using the approach developed in [4] let us write the expressions for $\tau_{SRH}$ value, when the recombination time determines the level of Fe for the SC base of p- and n- type, respectively, [5]:

$$\tau_{SRH}^p = \frac{\tau_{p0}(n_1+\Delta n)+\tau_{n0}(N_a+p_1+\Delta n)}{(N_a+\Delta n)}, \quad (1)$$

$$\tau_{SRH}^n = \frac{\tau_{p0}(N_d+n_1+\Delta n)+\tau_{n0}(p_1+\Delta n)}{(N_d+\Delta n)}, \quad (2)$$

where $\tau_{SRH}^p$ and $\tau_{SRH}^n$ are the SRH lifetimes for the p- and n-type bases, $\tau_{p0}=(1.19\cdot10^{-9}N_t)^{-1}$ s, $\tau_{n0}=(8.5\cdot10^{-7}N_t)^{-1}$ s, $n_1$ and $p_1$ are electrons and holes concentration for the cases when the position of the recombination level coincides with the Fermi level, and $n_1=n_i(T)\exp(-8.327)$, $p_1=n_i(T)\exp(8.327)$, $n_i(T)$ is the intrinsic carriers concentration in silicon, $T=25°C$. Numerical parameters for $\tau_{p0}$, $\tau_{n0}$, $n_1$ and $p_1$ calculation are taken from [3].

Hereinafter, the case of Si with high effective bulk lifetime values will be considered, when the minority charge carriers diffusion length is significantly higher than the SC base thickness $d$ throughout the doping level range, i.e. the effective diffusion length $L_{eff}=(D_{n(p)}\tau_{eff}^{n(p)})^{1/2} >> d$. Here $D_{n(p)}$ and $\tau_{eff}^{n(p)}$ are the diffusion coefficient and effective lifetime for electrons (holes). In this case, the excess minority charge carriers concentration $\Delta n$ is constant along the base.

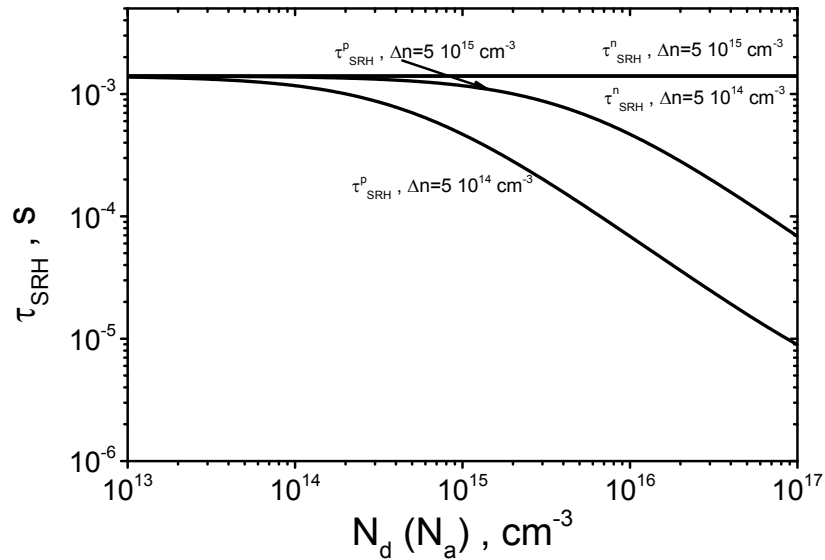

Fig.1. SRH lifetime versus the base doping level: curves 1 and 2 are for the p-type base and curves 3 and 4 are for the n-type base.

Fig.1 demonstrates dependence of $\tau_{SRH}^p$ and $\tau_{SRH}^n$ on the base doping level for two $\Delta n$ values: $5 \cdot 10^{14}$ cm$^{-3}$ and $5 \cdot 10^{15}$ cm$^{-3}$. Note that the first value for the base doping levels less or equal to $10^{15}$ cm$^{-3}$ is typical of the silicon solar cells operating at AM1.5 under the maximum power take-off conditions and the second value is typical of the case of the silicon solar cells operating in the open circuit mode. When calculating the $\tau_{SRH}$ value, the $N_t$ value was supposed to be equal to $6 \cdot 10^{11}$ cm$^{-3}$. Fig. 1 shows that the $\tau_{SRH}^p$ value decreases strongly (more than two orders of magnitude) with the base doping level increase, and strongly depends on the $\Delta n$ value. At the same time the $\tau_{SRH}^n$ value practically does not depend either on the base doping or on the $\Delta n$ value. With the concentration of the iron recombination centres equal to $6 \cdot 10^{11}$ cm$^{-3}$ chosen for calculations it is equal to $\tau_{SRH}^n$ = 1.4 ms. Such a lifetime is typical for high quality silicon.

## 2. Fundamental relations determining the silicon SC efficiency

Using the approach developed in [4] let us write the relations determining the photovoltaic conversion efficiency of the highly effective silicon SC and HIT SC η. Let us note in the beginning that in addition to the $L_{eff} \gg d$ inequality the $\Delta n_0 \geq N_d(N_a)$ condition is usually implemented, where $\Delta n_0$ is the excess electron-hole pairs concentration in the base for the open circuit case. So, the expressions for the open circuit voltage $V_{OC}$ in the p- and n-type bases case can be written, respectively, as [6]

$$V_{OC} \cong \frac{kT}{q}\ln\left(\frac{\Delta n_0}{p_0}\right) + \frac{kT}{q}\ln\left(1 + \frac{\Delta n_0}{N_d}\right), \qquad (3)$$

$$V_{OC} \cong \frac{kT}{q}\ln\left(\frac{\Delta n_0}{n_0}\right) + \frac{kT}{q}\ln\left(1 + \frac{\Delta n_0}{N_a}\right), \qquad (4)$$

where $k$ is the Boltzmann constant, $T$ is SC temperature, $q$ is the elementary charge, $p_0 = n_i^2(T)/N_d$ is the equilibrium concentration of holes in the n-type base, and $n_0 = n_i^2(T)/N_a$ is the equilibrium concentration of electrons in the p-type base.

For $\Delta n_0 \geq N_d(N_a)$ the open-circuit voltage $V_{OC}$ is higher than $V_{OC}$ for the standard case, when $\Delta n_0 < N_d(N_a)$.

The generation-recombination balance equation in the case of the open-circuit mode, when implementing inequality $L_{eff} \gg d$, can be written as

$$J_{SC}/q = \left[\frac{d}{\tau_b} + S\right]\Delta n_0, \qquad (5)$$

where $\tau_b = \left(\tau_{SRH}^{-1}(\Delta n_0) + A(N_d(N_a)) + \Delta n_0) + R_{Auger}\right)^{-1}$ is the bulk lifetime, $A \approx 6 \cdot 10^{-15}$ cm$^3$/s [7] is the radiative recombination coefficient in silicon, $S = S_0 + S_d$. The rate (inverse time) of the band-to-band Auger recombination ($R_{Auger}$) in the n - type silicon is determined by the expression

$$R_{Auger} = C_n(N_d + \Delta n_0)^2 + C_p(N_d + \Delta n_0)\Delta n_0, \qquad (6)$$

where $C_n = \left(2.8 \cdot 10^{-31} + \frac{2.5 \cdot 10^{-22}}{(N_d + \Delta n_0)^{0.5}}\right)$ cm$^6$/s, $C_p = 10^{-31}$ cm$^6$/s [8,9]. The second term in the expression for $C_n$ takes into account the many-electron effects, namely the effect of spatial correlation for distribution of two electrons and one hole involved in the act of Auger recombination, conditioned by the Coulomb interaction [8].

In the p-type silicon

$$R_{Auger} = C_p(N_a + \Delta n_0)^2 + C_n(N_a + \Delta n_0)\Delta n_0, \text{ and } C_n = \left(2.8 \cdot 10^{-31} + \frac{2.5 \cdot 10^{-22}}{\Delta n_0^{0.5}}\right) \text{ cm}^6/\text{s} . \quad (7)$$

Note that the expression for $C_n$ is the empirical one. A choice of this expression is discussed in more details in [9]. It was demonstrated that various dependencies were obtained in different experiments. The analysis of numerous experimental data, obtained in the case of silicon, makes it possible to derive the empirical expression fully describing the dependence of $R_{Auger}$ on the equilibrium concentrations of electrons $n_0$ and holes $p_0$ in the SC base, as well as on the concentration of excess electron-hole pairs $\Delta n$ at T=300 K

$$R_{Auger} = \frac{(n_0 + \Delta n)(p_0 + \Delta n)(1.8 \cdot 10^{-24} n_0^{0.65} + 6 \cdot 10^{-25} p_0^{0.65} + 3 \cdot 10^{-27} \Delta n^{0.8} + 9.5 \cdot 10^{-15})}{\Delta n}. \quad (8)$$

The last term in the rightmost parentheses (8) takes into account the radiative recombination.

Using expressions (5) and (6), the empirical expression like (8) can be derived for n-based SC substituting $6 \cdot 10^{-15}$ cm³/s for $9.5 \cdot 10^{-15}$ cm³/s. This expression agrees within less than six percent accuracy with (8) for $10^{15}$ cm⁻³ ≤ $n_0$ ≤ $4 \cdot 10^{16}$ cm⁻³. It can be written as

$$R_{Auger} = \left[(n_0 + \Delta n)\left(10^{-31}\Delta n + 2.8 \cdot 10^{-31}(n_0 + \Delta n) + 2.5 \cdot 10^{-22}(n_0 + \Delta n)^{0.5}\right) + 6 \cdot 10^{-15}\left(1 + 2.5 \cdot 10^{-16} n_0\right)\right] \quad (9)$$

As a result, the bulk lifetime $\tau_b$ values, calculated both using (8) and (9) in the actual for calculation of doping levels range, also differ less than by 6% (see Fig. 2).

Equation (5) is the quadratic equation in regard to the $\Delta n_0$ value and its solution can be written as

$$\Delta n_0 = -\frac{N_d(N_a)}{2} + \sqrt{\frac{(N_d(N_a))^2}{4} + n_i^2 \exp\left(\frac{qV_{OC}}{kT}\right)}. \quad (10)$$

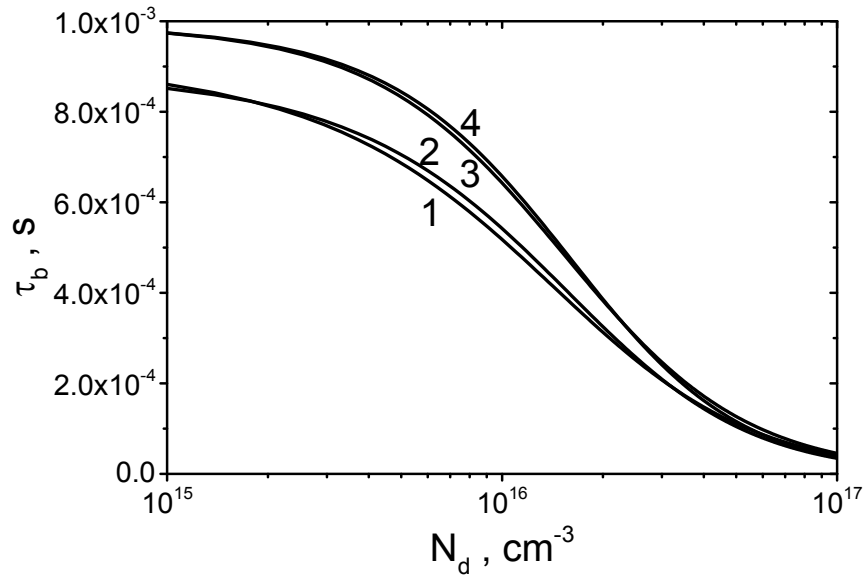

Fig.2. Dependence of the bulk base lifetime on the n-type base doping level of SC obtained with (8) (curves 1 and 3) and (9) (curves 2 and 4).

To calculate the photovoltaic conversion efficiency a theoretical expression for SC current-voltage characteristic is needed. With this aim in view, we proceed as follows. Let us replace $V_{OC}$ in (10) by the applied forward bias $V$ value. This operation makes it possible to determine $\Delta n(V)$

$$\Delta n(V) = -\frac{N_d(N_a)}{2} + \sqrt{\frac{(N_d(N_a))^2}{4} + n_i^2 \exp\left(\frac{qV}{kT}\right)}. \quad (11)$$

Next let us generalize equation (5), correct for the open circuit case, for the $V < V_{OC}$ case, i.e. for nonzero current. Then this equation can be written as:

$$J(V) = J_{SC} - J_{rec}(V), \quad (12)$$

where

$$J_{rec}(V) = q\left(\frac{d}{\tau_b} + S\right)\Delta n(V) \quad (13)$$

and the $\tau_b$ value can be found using (5) (with the replacement of $\Delta n_0$ by $\Delta n(V)$).

The $V_m$ value is found from the maximum power take-off condition $d(VJ(V))/dV = 0$ and its substitution in (13) makes it possible to determine the value of $J_m$. As a result, the photovoltaic conversion efficiency for the SC with unit area and series resistance $R_s$ per unit area is

$$\eta = \frac{J_m V_m}{P_S}\left(1 - \frac{J_m R_s}{V_m}\right), \quad (14)$$

where $P_S$ is the incident solar radiation power density.

### 3. Comparative analysis of the obtained relations for the SC base of n- and p- types

When building theoretical dependences for characteristics for the silicon SC with the base of the n-type the expressions for $R_{Auger}$ defined both by the relation (9) and the expression (8) were used. The following figures were built using (9) for the SC with the *n*-type base and (5), (7) in the case of the SC with *p*-type base.

Fig. 3 demonstrates the theoretical dependence for the open-circuit voltage on the base doping level for the p- and n-type base. The same parameters values determining the Shockley-Read-Hall lifetime were used in construction of the curves of Fig. 3 and Fig.1. The values of the surface recombination rate S and short-circuit current density $J_{SC}$ varied when calculating the curves 1 – 4. As Fig. 3 suggests, the $V_{OC}$ values coincide both for n- and p- type bases when the doping levels are low ($\leq 10^{15}$ cm$^{-3}$) (compare 1, 2 and 3, 4 curves calculated with the same $S$ and $J_{SC}$ values). It should be noted that $V_{OC}$ values do not depend on the doping level if $\Delta n_0 \gg N_d(N_a)$ inequality is satisfied. With $N_d \geq 10^{15}$ cm$^{-3}$ the $V_{OC}$ values in the SC with the n-type base increase, subsequently they reach a maximum and begin to decrease. In this case the $V_{OC}$ value decline is due to the prevalence of the Auger band-to-band recombination.

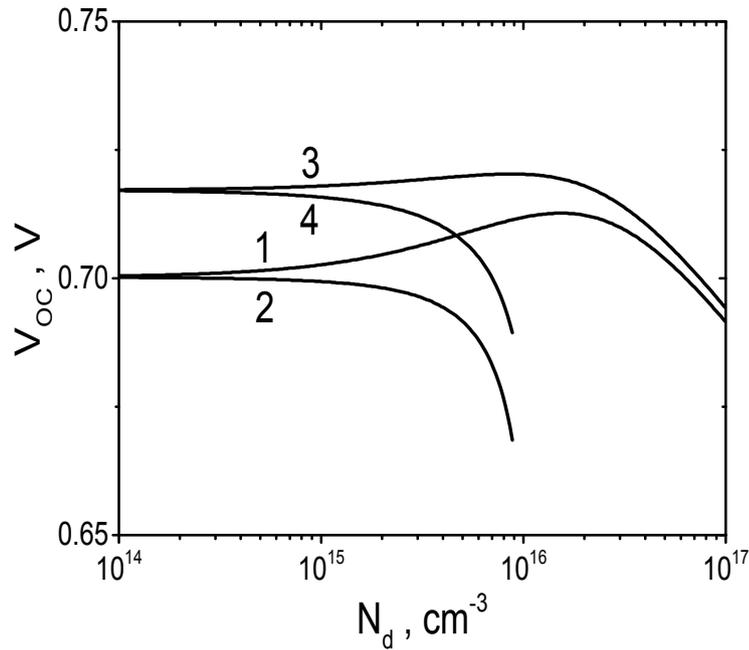

Fig.3. Dependences of the open circuit voltage on the n-type base doping level (curves 1 and 3) and p-type base doping level (curves 2 and 4). Parameters used for calculation: $d=300$ μm, $\tau_{SRH}=1.4$ ms, T=298 K. Curves 1 and 3 calculated with $S=12$ cm/s, $J_{SC}=36$ mA/cm$^2$ and curves 2 and 4 calculated with $S=1$ cm/s, $J_{SC}=39.5$ mA/cm$^2$.

The $V_{OC}(N_a)$ dependences in the SC with p-type base behave quite differently. With $N_a \geq 10^{14}$ cm$^{-3}$ the $V_{OC}$ values initially decrease slowly and with $N_a > 5 \cdot 10^{15}$ cm$^{-3}$ the rate of $V_{OC}$ decline increases significantly. In this case the $V_{OC}(N_a)$ decline is associated with the Shockley-Read-Hall lifetime decrease. Strictly speaking, in the framework of the used approximations the calculated dependences $V_{OC}(N_a)$ will be correct only with the fulfilment of the $L > 2d$ condition, which is implemented with $N_a < 9 \cdot 10^{15}$ cm$^{-3}$.

At higher doping levels not all of electron-hole pairs generated in the SC base will reach the p-n junction, that will primarily result in a short-circuit current reduction. But, the $V_{OC}$ values will also reduce as compared with those ones obtained in calculation in the assumption about the implementation of the $L >> d$ inequality.

Fig. 4 demonstrates the theoretical dependences of voltage $V_m$ on the p- and n- type base doping level in the mode of the maximum power take off. The same parameters values were used for construction of Fig.4 as for Fig. 3.

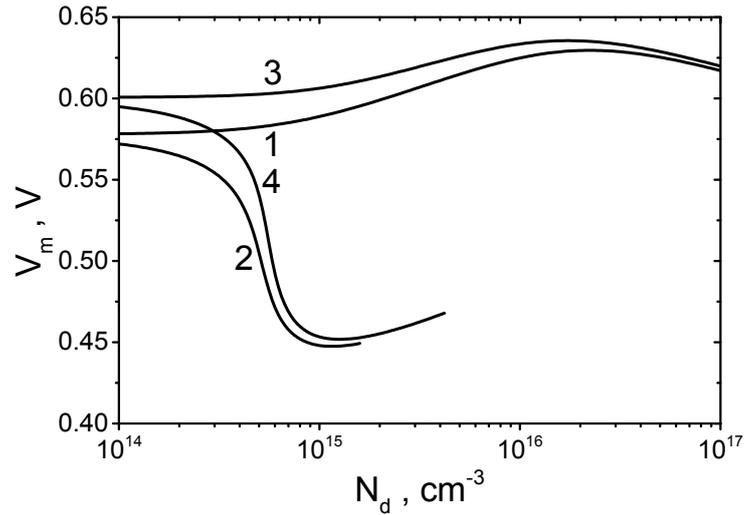

Fig.4. Dependence of voltage on the base doping level in the mode of the maximum power take off: curves 1 and 3 correspond to the n- type base, curves 2 and 4 correspond to the p-type base. $R_s = 0.21$ Ohm.

As is evidenced by the comparison between Figs. 3 and 4 the behaviour of $V_{OC}$ and $V_m$ depending on the base doping level is very similar in both cases with the only difference that for the dependences $V_m(N_d(N_a))$ the $V_m$ value region of independence on $N_d(N_a)$ and the point of the $V_m(N_a)$ minimum are shifted to the region of lower doping levels' values. Thus, the region of independence of $V_m$ on $N_d(N_a)$ is realized when $N_d(N_a) \leq 10^{14}$ cm$^{-3}$, and the point of the $V_m$ minimum is realized when $N_a \approx 10^{15}$ cm$^{-3}$. It should be noted that in case of the p-type base the $L > 2d$ condition, necessary for the $V_m$ correct calculation, is fulfilled if $N_a \leq 10^{15}$ cm$^{-3}$.

Fig. 5 demonstrates the theoretical dependences of the photovoltaic conversion efficiency $\eta$ on the doping level of the p- and n-type bases. The same parameter values were used for construction of Fig. 5 as for Figs. 3 and 4. As Fig. 5 suggests, the dependences of photovoltaic conversion efficiency $\eta$ on the base doping level repeated dependence $V_m$ on $N_d(N_a)$ (see Fig. 4) with a certain scaling. The essential difference of these dependences for the SC with the base of the n- and p- types is that in the SC with the n- type base the $\eta$ values grow with the growth of the base doping level and in solar cells with p- type base the $\eta$ values decrease. As already mentioned above, the decrease in $\eta(N_a)$ with the increase in $N_a$ in the SC with the p - type base is associated with a decrease in the Shockley-Read-Hall lifetime. Fig. 5 shows that for the typical doping levels of SC bases ($\sim (2-4) \cdot 10^{15}$ cm$^{-3}$) the efficiency of solar cells with the n-type base significantly exceeds the efficiency of solar cells with the p - type base. Increase in the

$\eta$ value in the SC with the p - type base can be achieved by reducing the concentration of the recombination centres of iron.

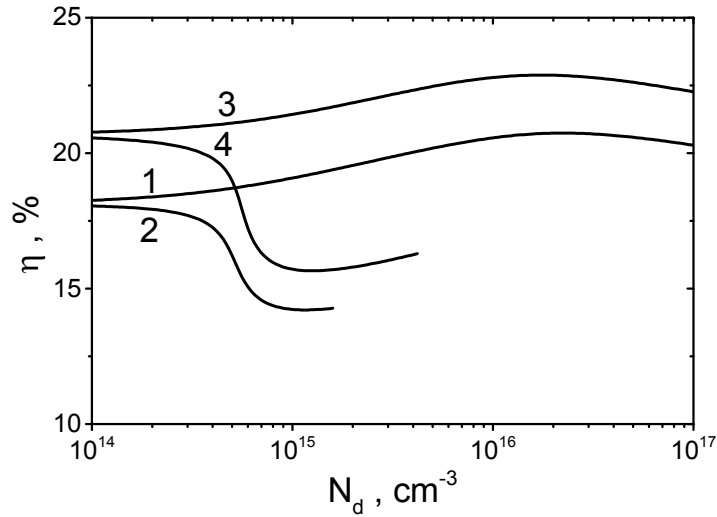

Fig.5. Dependence of the photovoltaic conversion efficiency on the doping level base: curves 1 and 3 correspond to the n- type base and curves 2 and 4 correspond to the p-type base. The same parameter values were used for construction of Fig.5 as for Fig. 3. $R_s$ =0.21 Ohm.

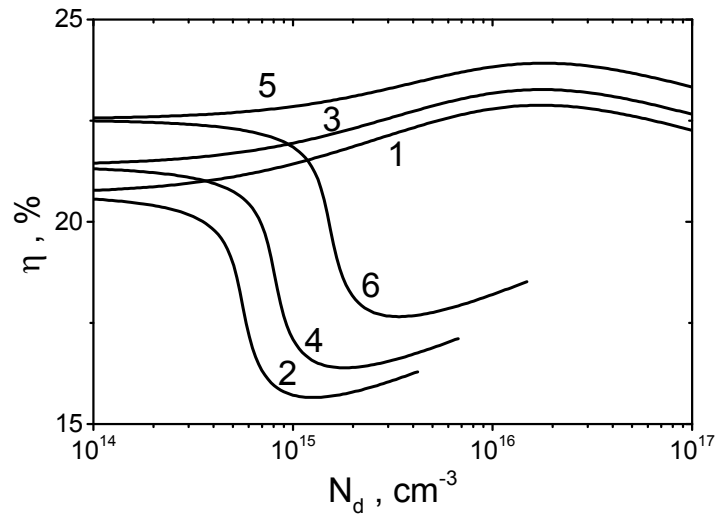

Fig.6. Photovoltaic conversion efficiency dependence on the base doping level $N_d(N_a)$ with different base thicknesses: curves 1, 3, 5 are plotted for the n- type base and curves 2, 4, 6 are plotted for the n- type base. The following parameters were used in construction of the plot: $J_{SC}$ = 39.5 mA/cm$^2$, $S$ =1 cm/s, $R_s$ = 0.21 Ohm. The base thickness $d$ was equal to: 1, 2 -300 μm, 3, 4 – 200 μm, 5, 6 -100 μm. The rest parameters were the same as those used in Fig. 3 curves construction.

Fig. 6 shows the calculated dependence of the photovoltaic efficiency $\eta$ on the base doping level for the n- and p- type bases for the case where only the base thickness varies. The calculation was performed for the case when the base thickness $d$ was equal to 100, 200 and 300 microns. As can be seen from Fig. 6 the lower the thickness of the base, the greater the efficiency $\eta$. This increase is associated with the increase in the open circuit voltage $V_{OC}$, more precisely, in the voltage in the mode of the maximum power take off $V_m$ due to the bulk recombination reduction.

The doping levels' values, with which the dependences $\eta(N_a)$ are broken in case of the SC with the p-type base, correspond to the $L_{eff}(N_a) \approx 2d$ conditions. As Fig. 6 suggests, the lower the base thickness $d$ the higher doping levels required for the $L_{eff}(N_a) \geq 2d$ condition fulfilment.

## 4. Comparison of the experimental and calculated $\alpha-Si:H-n-Si$ HIT cells parameters' values

Let us compare next the calculated and experimental characteristics of HIT solar cells with those ones obtained in the experiment in the SC with the base of n – type: $J_{SC}$=36.03 mA/cm$^2$, $V_{OC}$=0.703 V, FF=0.77, $\eta$= 19.4 %, $S_{SC}$=3.96 cm$^2$ for the case when the SC base region parameters were equal to the following values: $N_d \approx 1.6 \cdot 10^{15}$ cm$^{-3}$, $d \approx 300$ μm, $\tau_{SR} \approx 1.4$ ms.

HIT solar cell manufacturing technology included operations of cleaning of crystalline silicon wafer surface, surfaces texturing (pyramidal surface relief patterning), acid underetching of the surface layer, putting of optimum thickness layer of intrinsic amorphous hydrogenated silicon $(i)\alpha-Si:H$ on both wafer surfaces. Then $n-cSi/(n^+)\alpha-Si:H$ isotype heterojunction was formed on the back side surface and $n-cSi/(p^+)\alpha-Si:H$ anisotype heterojunction was formed on the front side surface. The transparent conductive layers of indium and tin oxides mixture were deposited on both surfaces and then the low-temperature annealing was realized to reduce the series resistance. The contacts deposition was the finishing operation: a solid contact was deposited on the back side and the netlike contact was deposited on the front side. The HIT solar cell area was about 4 cm$^2$.

To calculate $R_{Auger}$ the relation (9) was used when preparing Table 1 and the relation (8) was used when preparing Table 2, respectively. The first line of Tables 1 and 2 shows the calculated characteristics of the HIT solar cells and base region parameters obtained as a result of calculation by given above formulas. As can be seen from a comparison of the calculated and experimental parameters given in the first lines of Tables 1 and 2, they are identical, if $S$ =12 cm/s, and $R_s$ =0.21 Ohm. Note that the calculated value of the fill factor $FF$ for current-voltage characteristic in this case is 0.77 and it coincides well with the experimental value.

Table 1

| Sample | $N_d$, cm$^{-3}$ | d, µm | $\tau_{SRH}$, ms | $J_{SC}^{exp}$, mA/cm$^2$ | $V_{OC}$, V | $S$, cm/s | $\eta$, % | $R_s$, Ohm |
|---|---|---|---|---|---|---|---|---|
| n- type | $1.6 \cdot 10^{15}$ | 300 | 1.4 | 36.03 | 0.704 | 12 | 19.4 | 0.105 |
| n- type | $1.6 \cdot 10^{15}$ | 300 | 1.4 | 39.5 | 0.719 | 1 | 21.8 | 0.105 |
| n- type | $4 \cdot 10^{16}$ | 300 | 1.4 | 36.03 | 0.709 | 12 | 21.0 | 0.105 |
| n- type | $2 \cdot 10^{16}$ | 300 | 1.4 | 39.5 | 0.719 | 1 | 23.2 | 0.105 |

Table 2

| Sample | $N_d$, cm$^{-3}$ | d, µm | $\tau_{SRH}$, ms | $J_{SC}^{exp}$, mA/cm$^2$ | $V_{OC}$, V | $S$, cm/s | $\eta$, % | $R_s$, Ohm |
|---|---|---|---|---|---|---|---|---|
| n- type | $1.6 \cdot 10^{15}$ | 300 | 1.4 | 36.03 | 0,704 | 12 | 19.4 | 0.105 |
| n- type | $1.6 \cdot 10^{15}$ | 300 | 1.4 | 39.5 | 0,719 | 1 | 21.7 | 0.105 |
| n- type | $4 \cdot 10^{16}$ | 300 | 1.4 | 36.03 | 0,710 | 12 | 20.7 | 0.105 |
| n- type | $2 \cdot 10^{16}$ | 300 | 1.4 | 39.5 | 0,719 | 1 | 22.9 | 0.105 |

For the moderate levels of the base doping (~$1.6 \cdot 10^{15}$ cm$^{-3}$) the HIT solar cell parameters, given in the second line of Tables 1 and 2, practically coincide. The calculated SC parameters were obtained using the following values: $S$=1 cm/s, $J_{SC}$=39.5 mA/cm$^2$ [4, 11].

The third line of Table 1 and Table 2 presents the calculated $V_{OC}$ and $\eta$ values for the case when the base doping level is $4 \cdot 10^{16}$ cm$^{-3}$, $S$=12 cm/s, $J_{SC}$=36 mA/cm$^2$. At this doping level the $\eta$ value is the peak value. As can be seen from the tables, the $V_{OC}$ and $\eta$ values for this case increase as compared with the case where the doping level of the base is $1.6 \cdot 10^{15}$ cm$^{-3}$. $V_{OC}$ and $\eta$ values are somewhat higher, when the expression (6) is used for $R_{Auger}$. The discrepancy between the values of $\eta$ in this case is about 1% and the discrepancy between the $V_{OC}$ values is virtually absent.

The calculated HIT SC characteristics, given in the fourth line of Tables 1 and 2, are also somewhat different when built with $S$=1 cm/s and $J_{SC}$=39.5 mA/cm$^2$. In this case, the maximum values of $V_{OC}$ and $\eta$ are realized at the doping level of $2 \cdot 10^{16}$ cm$^{-3}$. In this case the $V_{OC}$ and $\eta$ values are the peak values.

Thus, as can be seen from the above analysis, the base doping level is the parameter which should be optimized to achieve the maximum value of $\eta$ when the rest parameters are set.

## 5. Conclusions

The theoretical analysis of high-efficient silicon SC efficiency is carried out depending on the type and level of doping of the base for the case when Fe level induces the SRH recombination. It is shown that the photovoltaic conversion efficiency in case of the n-type base increases with the increase in the level of the base doping level, reaches its maximum and begins to decrease. This decrease is determined by the band-to-band Auger recombination predominance. In the case of the SC with the p-type base, the photovoltaic conversion efficiency decreases with base doping level increase. The obtained result is associated with a strong asymmetry of Fe level parameters, due to it a strong decrease in the $\tau_{SRH}$ value occurs in the p-type samples in the field of the doping levels' values range characteristic for the Si solar cells.

The calculated and experimental values of characteristics of the HIT SC based on $\alpha-Si:H-n-Si$ are compared. The effective surface recombination velocity and serial resistance values are defined from the comparison; complete coincidence is achieved between theory and practice.

It is shown that the base doping level is one of the most important parameters of the HIT solar cells. This parameter is subject to optimization and the use of its optimal values for the SC of the order of $(2 - 4) \cdot 10^{16}$ cm$^{-3}$ with the n-type base makes it possible to increase significantly the $\eta$ value.